\documentclass[cameraready]{Interspeech}

\title{NVV-SuperBench: Beyond Words, Beyond Quality---Benchmarking Nonverbal Vocalizations in Speech Generation}
\author[affiliation={1,2}]{Liumeng}{Xue}
\author[affiliation={2}]{Weizhen}{Bian}
\author[affiliation={2}]{Jiahao}{Pan}
\author[affiliation={3}]{Wenxuan}{Wu}
\author[affiliation={4}]{Yilin}{Ren}
\author[affiliation={2}]{Boyi}{Kang}
\author[affiliation={5}]{Jingbin}{Hu}
\author[affiliation={6}]{Ziyang}{Ma}
\author[affiliation={1}]{Shuai}{Wang}
\author[affiliation={4}]{Xinyuan}{Qian}
\author[affiliation={7}]{Hung-yi}{Lee}
\author[affiliation={2}]{Yike}{Guo}

\address{
$^1$ Nanjing University \\
$^2$ The Hong Kong University of Science and Technology \\
$^3$ The Chinese University of Hong Kong \\
$^4$ University of Science and Technology Beijing \\
$^5$ Northwestern Polytechnical University \\
$^6$ Shanghai Jiao Tong University \\
$^7$ National Taiwan University
}

\email{lmxue@nju.edu.cn}


\usepackage{comment}

\newcommand{\g}{\cellcolor{gray!12}}

\usepackage{booktabs}
\usepackage{hyperref}
\usepackage{url}
\usepackage{xurl} 
\usepackage{tikz}
\usetikzlibrary{arrows.meta, shapes, positioning, calc}
\usepackage{multirow}
\usepackage{xcolor}
\usepackage{amsmath,amssymb}
\usepackage{graphicx}
\usepackage{makecell}
\usepackage[table]{xcolor} 
\usepackage{subcaption} 
\usepackage{tabularx}
\usepackage{array}
\usepackage{float}
\usepackage{listings}
\usepackage{xcolor} 
\usepackage[T1]{fontenc}
\usepackage{textcomp}
\usepackage{adjustbox}
\usepackage{caption}
\usepackage[font=small,skip=2pt]{caption} 
\lstdefinestyle{jsoncompact}{
  basicstyle=\ttfamily\footnotesize,
  breaklines=true,
  columns=fullflexible,
  keepspaces=true,
  showstringspaces=false,
  frame=single,
  upquote=true,
  literate={%
    {"text"}{{\textbf{"text"}}}{6}%
    {"text_with_nvv"}{{\textbf{"text\_with\_nvv"}}}{15}%
    {"caption_with_nvv"}{{\textbf{"caption\_with\_nvv"}}}{23}%
    {"nvv_list"}{{\textbf{"nvv\_list"}}}{21}%
  },
}

\begin{document}

\maketitle

\begin{abstract}
\vspace{-4pt}
Nonverbal vocalizations (NVVs), such as laughing, sighing, and sobbing, are essential for human-like speech, yet standardized evaluation rarely jointly assesses whether systems generate the intended NVVs, place them correctly, and keep them salient without harming speech. We present \textbf{NVV-SuperBench}, a bilingual English/Chinese benchmark for speech generation with NVVs. It provides a unified 45-type taxonomy and a multi-axis protocol beyond conventional speech quality assessment, evaluating NVV-specific controllability, placement, and perceptual salience. We benchmark 15 speech generation systems spanning prompt-based and tag-based control paradigms, using objective metrics, human listening tests, and LLM-based multi-rater evaluation. Results show that NVV controllability often decouples from speech quality, while low-SNR oral cues and long-duration affective NVVs remain bottlenecks. NVV-SuperBench highlights current gaps and supports progress toward more human-like speech generation.

\end{abstract}

\vspace{-6pt}
\section{Introduction}
\vspace{-4pt}
Speech generation has progressed rapidly, moving beyond intelligible speech toward expressive and controllable generation.
Recent large-scale speech language modeling and codec-based generation paradigms have further improved perceptual quality, speaker similarity, and controllability, pushing synthetic speech toward increasingly human-like delivery~\cite{cui2025speechlm_survey,spark_tts2025,voxcpm_iclr2026}.  However, achieving truly human-like spoken interaction requires more than just accurate lexical content. It necessitates the inclusion of nonverbal vocalizations (NVVs), such as laughter, sighs, and gasps, which carry essential affective and social signals that are critical for emotional communication and immersive human-computer interaction.

\begin{table}[th]
\centering
\caption{45-type NVV taxonomy of NVV-SuperBench.}
\scriptsize
\setlength{\tabcolsep}{3pt}
\renewcommand{\arraystretch}{1.05}
\scalebox{0.93}{  
\begin{tabularx}{\linewidth}{@{}p{0.28\linewidth}Xc@{}}
\toprule
\textbf{Category} & \textbf{NVV type} & \textbf{Count} \\
\midrule
\textbf{Respiratory} &
breath, inhale, exhale, quick breath, sigh, gasp, panting, wheezing, snore, yawn & 10 \\
\textbf{Throat / Physiological} &
cough, sneeze, throat clearing, hiccup, sniff, sniffle, snort & 7 \\
\textbf{Laughter Spectrum} &
chuckle, giggle, laugh, laugh harder, start laughing, stifled laugh, burst of laughter & 7 \\
\textbf{Crying Spectrum} &
crying, sobbing, crying loudly, wail, whimper & 5 \\
\textbf{Emotional Vocalizations} &
hum, humming, groan, moan, grunt, mumble, exclamation (ah, oh, hmm) & 7 \\
\textbf{Oral / Miscellaneous} &
lipsmack, gulp, swallow, burp, tsk, sss, clucking, hissing, whisper & 9 \\
\midrule
\textbf{Total} &  & 45 \\
\bottomrule
\end{tabularx}
}
\label{table:nvv_taxonomy}
\vspace{-5pt}
\end{table}

Nonverbal vocalizations are crucial in intention understanding, emotional communication, and immersive human--computer interaction by conveying emotional and social context that words alone cannot express~\cite{eyben2016gemaps}. 
These vocalizations encompass a wide range of behaviors, from discrete events like laughter or sighs, to low-energy oral cues such as lip smacks, tongue clicks, and breathy sounds, which are often masked by surrounding phonation. Additionally, long-duration affective vocal styles, such as crying, trembling speech, and breathy exhaustion, contribute to the tone and intent of an utterance, requiring consistent modulation of prosody, timbre, and voice quality over time.

Despite their importance, synthesizing NVVs remains a complex task. These vocalizations need to be seamlessly integrated into natural speech while preserving their emotional and social significance. NVV synthesis involves not only accurate sound representation but also ensuring the correct emotional tone and contextual appropriateness, especially in long-duration expressions. 
Recent research has explored NVV-related recognition, data construction, and generation~\cite{borisov2025nonverbaltss,nonverbalspeech38k2025,wu2025smiipnv,liao2025nvspeech,mnv17_2025}, and has also established benchmarks for NVV detection and localization~\cite{wesr2026}. However, a standardized and comprehensive framework for evaluating NVV synthesis in full-sentence speech generation is still lacking, making it difficult to assess and improve speech generation systems' capabilities in this area.

Speech generation evaluation efforts have moved beyond conventional assessments of speech quality to probe paralinguistic aspects of speech~\cite{paralbench2024,instructttseval2025,s2sarena2025,paras2s2025,wavbench2026,chen2026mint}. Recent work has also begun to benchmark nonverbal vocalization synthesis in speech generation~\cite{ni2026nvbench}.
However, fine-grained NVV coverage, bilingual evaluation, diverse control interfaces, and NVV-specific attributes such as controllability, placement, and salience remain underexplored.

\begin{figure*}[!th]
    \centering
    \includegraphics[width=0.72\linewidth]{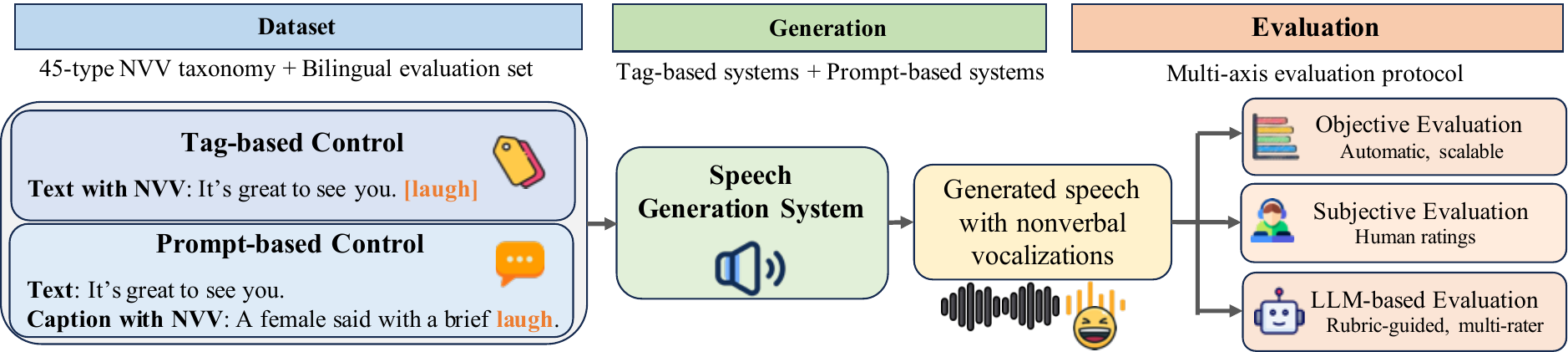}
    \caption{Overview of NVV-SuperBench.}
    \label{fig:overview_nvbench}
    \vspace{-8pt}
\end{figure*}

To address the gap, we introduce \textbf{NVV-SuperBench}, a bilingual (English/Chinese) benchmark for evaluating \emph{speech generation with nonverbal vocalizations}\footnote{Available at \href{https://lmxue.github.io/NVV-SuperBench/}{https://lmxue.github.io/NVV-SuperBench/}}.
NVV-SuperBench covers a unified 45-type NVV taxonomy with a curated bilingual evaluation set, and introduces a multi-axis evaluation protocol that disentangles general speech naturalness and quality from NVV-specific \emph{controllability}, \emph{placement}, and \emph{perceptual salience}. We benchmark 15 representative speech generation systems (8 tag-based systems and 7 prompt-based systems), measured via objective metrics, human listening tests, and LLM-based multi-rater evaluation. 
Results indicate that NVV controllability often decouples from overall speech quality and differs markedly across control interfaces, and highlight low-SNR oral cues and long-duration affective NVVs as persistent bottlenecks.
By grounding all systems in the same taxonomy, data, and evaluation criteria, NVV-SuperBench facilitates fair cross-system comparisons across diverse control interfaces. In summary, our contributions are:

\begin{itemize}
  \item \textbf{Taxonomy \& benchmark set.} We introduce a unified 45-type NVV taxonomy and curate a bilingual (English/Chinese) evaluation set to enable consistent, cross-lingual assessment of speech generation with NVVs.
  \item \textbf{Multi-axis evaluation.} We propose a multi-axis evaluation protocol that disentangles general speech naturalness and quality from NVV-specific \emph{controllability}, \emph{placement}, and \emph{perceptual salience} with objective metrics, human listening tests, and LLM-based multi-rater evaluation.
  \item \textbf{Comprehensive system study.} We benchmark 15 representative speech systems, spanning commercial and open-source models, and reveal that low-SNR oral cues and long-duration affective NVVs remain persistent bottlenecks, motivating richer NVV inventories and improved coherence for sustained NVVs.

\end{itemize}

\newcommand{\nvvsep}{,\allowbreak\hspace{0.15em}}   
\newcommand{\nvvsepD}{,\allowbreak\hspace{0.45em}}  
\setlength{\aboverulesep}{0pt}
\setlength{\belowrulesep}{0pt}

\begin{table*}[ht]
\centering
\caption{NVV inventories of representative tag-based speech generation systems and datasets. 
}
\scriptsize
\setlength{\tabcolsep}{1pt}
\renewcommand{\arraystretch}{1.05}
\begingroup
\renewcommand{\tabularxcolumn}[1]{>{\raggedright\arraybackslash}m{#1}}
\scalebox{0.85}{
\begin{tabularx}{\textwidth}{@{}
>{\centering\arraybackslash}m{0.075\textwidth}   
>{\raggedright\arraybackslash}m{0.15\textwidth}  
X                                                
>{\raggedleft\arraybackslash}m{0.04\textwidth}   
>{\centering\arraybackslash}m{0.06\textwidth}    
@{}}
\toprule
\textbf{Type} & \textbf{System / Dataset} & \textbf{Supported NVV Types} & \textbf{Count} & \textbf{Lang.} \\
\midrule

\multirow{8}{*}{\textbf{System}}
& ChatTTS~\cite{chattts_github_2024}
& {\ttfamily laugh}
& 1 & EN, ZH \\

& Higgs-Audio~\cite{higgsaudio_github_2025}
& {\ttfamily laugh\nvvsepD Humming\nvvsepD cough}
& 3 & EN, ZH \\

& Bark~\cite{bark_github_2023}
& {\ttfamily laughter\nvvsepD laughs\nvvsepD sighs\nvvsepD gasps\nvvsepD clears throat}
& 5 & EN, ZH \\

& Fish-Speech~\cite{liao2024fishspeech}
& {\ttfamily laughing\nvvsepD chuckling\nvvsepD sobbing\nvvsepD crying loudly\textsuperscript{\dag}\nvvsepD sighing\nvvsepD panting\nvvsepD groaning}
& 7 & EN, ZH \\

& Orpheus TTS~\cite{orpheus_tts_github_2025}
& {\ttfamily laugh\nvvsepD chuckle\nvvsepD sigh\nvvsepD cough\nvvsepD sniffle\nvvsepD groan\nvvsepD yawn\nvvsepD gasp}
& 8 & EN, ZH \\

& CosyVoice 2~\cite{du2024cosyvoice2}
& {\ttfamily breath\nvvsepD laughter\nvvsepD cough\nvvsepD clucking\nvvsepD quick\_breath\textsuperscript{\dag}\nvvsepD hissing\nvvsepD sigh\nvvsepD lipsmack}
& 8 & EN, ZH \\

& ElevenLabs~\cite{elevenlabs_models_docs}\textsuperscript{\S}
& {\ttfamily laughs\nvvsepD laughs harder\textsuperscript{\dag}\nvvsepD starts laughing\nvvsepD wheezing\nvvsepD whispers\nvvsepD sighs\nvvsepD exhales\nvvsepD crying\nvvsepD snorts\nvvsepD giggles\nvvsepD swallows\nvvsepD gulps}
& 12 & EN, ZH \\

& Dia~\cite{dia_github}
& {\ttfamily laughs\nvvsepD clears throat\nvvsepD sighs\nvvsepD gasps\nvvsepD coughs\nvvsepD groans\nvvsepD sniffs\nvvsepD inhales\nvvsepD exhales\nvvsepD burps\nvvsepD humming\nvvsepD sneezes\nvvsepD chuckle}
& 13 & EN \\

\midrule

\multirow{6}{*}{\textbf{Dataset}}
& SMIIP-NV~\cite{wu2025smiipnv}
& {\ttfamily laughter\nvvsepD crying\nvvsepD cough}
& 3 & ZH \\

& NVSpeech~\cite{liao2025nvspeech}
& {\ttfamily breath\nvvsepD crying\nvvsepD laughter\nvvsepD cough\nvvsepD sigh}
& 5 & ZH \\

& SynParaSpeech~\cite{bai2025synparaspeech}
& {\ttfamily sigh\nvvsepD throat\ clearing\nvvsepD laugh\nvvsepD tsk\nvvsepD gasp}
& 5 & ZH \\

& NonverbalTTS~\cite{borisov2025nonverbaltss}
& {\ttfamily breath\nvvsepD laugh\nvvsepD sniff\nvvsepD cough\nvvsepD throat\nvvsepD sigh\nvvsepD groan\nvvsepD sneeze\nvvsepD snore\nvvsepD grunt}
& 10 & EN \\

& NonverbalSpeech-38k~\cite{nonverbalspeech38k2025}
& {\ttfamily laughing\nvvsepD coughing\nvvsepD breath\nvvsepD sniff\nvvsepD crying\nvvsepD throat\ clearing\nvvsepD sigh\nvvsepD snore\nvvsepD gasp\nvvsepD yawn}
& 10 & EN, ZH \\

& MNV-17~\cite{mnv17_2025}
& {\ttfamily sighing\nvvsepD sneezing\nvvsepD clapping\nvvsepD hissing\nvvsepD whistling\nvvsepD clearing\ throat\nvvsepD coughing\nvvsepD lip\ smacking\nvvsepD exhaling\nvvsepD moaning\nvvsepD panting\nvvsepD sniffling\nvvsepD humming\nvvsepD laughing\nvvsepD applauding\nvvsepD inhaling\nvvsepD chuckling}
& 17 & ZH \\



\bottomrule
\end{tabularx}
}
\endgroup

\vspace{2pt}
\begin{minipage}{\textwidth}
\scriptsize\emph{\raggedright
Note:
\textsuperscript{\S} indicates commercial TTS systems.
\textsuperscript{\dag} marks tags with higher intensity, loudness, or speed. 
We exclude tags that do not correspond to nonverbal vocalizations (e.g., nonvocal action/sound-effect tags like \texttt{[clapping]} and purely stylistic tags like \texttt{[sarcastic]}) from the systems.}

\end{minipage}

\label{tab:nvv_tags_and_datasets}
\vspace{-5pt}
\end{table*}

\vspace{-9pt}
\section{NVV-SuperBench}
\vspace{-4pt}
\subsection{Benchmark overview}
\vspace{-4pt}
We introduce \textbf{NVV-SuperBench}, a standardized evaluation suite for assessing a speech generation system's ability to synthesize \emph{nonverbal vocalizations (NVVs)} beyond lexical content. The overview of NVV-SuperBench is presented in Figure~\ref{fig:overview_nvbench}.
Given an input utterance, NVV-SuperBench supports two commonly used NVV-control interfaces: 
(i) \textit{prompt-based} control, where NVV cues are specified in a natural-language caption; and 
(ii) \textit{tag-based} control, where NVV tags (e.g., \texttt{[laugh]}, \texttt{[sigh]}) are inserted into text.
A candidate speech generation system generates speech conditioned on the input, and NVV-SuperBench evaluates the speech output from three complementary perspectives:
automatic \textit{objective} metrics, human \textit{subjective} listening tests, and \textit{LLM-based multi-rater} evaluation.

\vspace{-6pt}
\subsection{NVV taxonomy}
\label{subsec:nvb-taxonomy}
\vspace{-4pt}
Nonverbal vocalizations (NVVs) are essential to human spoken communication, conveying physiological states (e.g., coughing and breathing), affective expressions (e.g., laughter and crying), and interactional signals (e.g., sighs, grunts, and whispers). These NVVs enhance conversational realism and emotional expressiveness, which are critical for achieving more human-like speech generation~\cite{liao2025nvspeech,wang2025capspeech}. To systematically benchmark speech generation systems beyond lexical content, we design a structured taxonomy of NVVs, as shown in Table~\ref{table:nvv_taxonomy}.

Specifically, we start by surveying the NVV inventories supported by representative commercial and open-source tag-based speech generation systems~\cite{chattts_github_2024,higgsaudio_github_2025,bark_github_2023,liao2024fishspeech,orpheus_tts_github_2025,du2024cosyvoice2,elevenlabs_models_docs,dia_github}, and by compiling the NVV labels reported in recent datasets~\cite{borisov2025nonverbaltss,nonverbalspeech38k2025,wu2025smiipnv,liao2025nvspeech,mnv17_2025}, as shown in Table~\ref{tab:nvv_tags_and_datasets}.
We also note recent benchmark efforts such as WESR~\cite{wesr2026} and NV-Bench~\cite{ni2026nvbench}. Their NVV types are largely covered by our unified taxonomy, although WESR primarily targets event-speech recognition rather than speech generation.
This survey shows that existing systems and datasets cover a limited, highly skewed subset of NVVs, with fragmented and sometimes inconsistent labels. 
Instead, we define a broader, model-agnostic NVV space to systematically probe system boundaries and generalization in synthesizing diverse nonverbal behaviors. 
Guided by production mechanisms and communicative functions, we therefore design a comprehensive NVV taxonomy with six top-level categories and 45 fine-grained types:

\vspace{-2pt}
\begin{itemize}
\setlength{\itemsep}{0pt}\setlength{\parskip}{0pt}\setlength{\topsep}{2pt}

  \item \noindent\textbf{Respiratory.}  
  Respiratory NVVs (e.g., inhalation, sigh) represent subtle \emph{breathing patterns} that are essential for controlling \emph{emotion intensity} in speech. Proper breath management is crucial for achieving natural \emph{emotion modulation}, a core component of \emph{speech generation}, making speech \emph{more human-like} in emotional expressiveness~\cite{werner2024acoustics,kamiloglu2024voiceswithoutwords}.

  \item \noindent\textbf{Throat / Physiological.}  
  This category encompasses reflexive vocalizations (e.g., cough, sniff), which are \emph{semi-voluntary sounds} central to real-world interactions. These cues enable the system to capture important \emph{physiological responses}, such as \emph{hesitation, discomfort}, or \emph{attention}, contributing to \emph{authentic conversational speech} where \emph{real-time emotional reactions} are integral~\cite{schuller2022compare_mm}.

  \item \noindent\textbf{Laughter spectrum.}  
  Laughter subtypes (e.g., chuckle, giggle, laugh) vary significantly in \emph{intensity}, \emph{duration}, and \emph{social role}, and must be synthesized accurately to replicate \emph{human emotional interactions}. By distinguishing different laughter types, we enable the system to produce more \emph{engaging}, \emph{empathetic}, and \emph{emotionally rich speech}, which is essential for \emph{more human-like speech generation} in conversational contexts~\cite{wang24b_speechprosody,ludusan24_speechprosody}.

  \item \noindent\textbf{Crying spectrum.}  
  Cry-related NVVs (e.g., sobbing, wailing) convey varying levels of \emph{distress} and \emph{vulnerability}, which are key to generating \emph{emotionally responsive systems}. Accurate generation of \emph{crying sounds} allows the system to exhibit \emph{genuine emotional depth}, enabling \emph{empathy-driven interactions} and improving \emph{emotional realism} in human-machine conversations~\cite{kamiloglu2024voiceswithoutwords}.

  \item \noindent\textbf{Emotional vocalizations.}  
  Non-lexical affective sounds (e.g., hum, moan, grunt) provide critical \emph{attitudinal context}, signaling \emph{emotion} or \emph{hesitation} in speech. These sounds help enhance the \emph{emotion dynamics} of the generated speech, making the conversation feel more \emph{authentic} and \emph{emotionally nuanced}, which is essential for achieving \emph{more human-like dialogue} in expressive speech generation~\cite{kamiloglu2024voiceswithoutwords,schuller2022compare_mm}.

  \item \noindent\textbf{Oral / Miscellaneous.}  
  Oral cues such as \emph{lipsmack}, \emph{hissing}, and \emph{whisper-like sounds} are key to achieving natural \emph{interactional speech}, especially for \emph{nonverbal exchanges}. These sounds help bridge gaps between \emph{spoken words}, enhancing \emph{fluidity} and \emph{responsiveness} in real-time conversations, which are essential for \emph{human-like conversational fluency} in speech generation systems~\cite{schuller2022compare_mm}.
\end{itemize}

To balance coverage and diagnostic clarity, we introduce fine-grained subtypes only when the variation is perceptually salient and is also commonly distinguished in prior datasets or speech generation control interfaces~\cite{borisov2025nonverbaltss,nonverbalspeech38k2025,liao2025nvspeech,wu2025smiipnv}.
For instance, we subdivide laughter to capture robust differences in intensity and conversational function that are reflected in acoustic--prosodic patterns~\cite{wang24b_speechprosody,ludusan24_speechprosody}, while we keep many physiological reflexes (e.g., cough and sneeze) at a coarser granularity to reduce sparsity and maintain cross-dataset consistency.
This design facilitates systematic evaluation beyond lexical accuracy and is consistent with recent efforts on NVV-aware speech generation research.

\vspace{-7pt}
\subsection{Data construction}
\label{subsec:nvb-data}
\vspace{-4pt}
To construct a type-balanced NVV dataset in both English and Chinese, we develop a three-stage pipeline that integrates (i) LLM-assisted seed mining from expressive human speech, (ii) taxonomy-driven controlled generation, and (iii) iterative validation with replenishment. This design yields a quality-controlled dataset with balanced per-type coverage across the NVV taxonomy. Example inputs are shown in Figure~\ref{fig:overview_nvbench}.

\textbf{Stage I: Seed mining from human speech.}
We begin by mining high-confidence NVV \emph{seeds} from \emph{human} speech, rather than directly synthesizing data from scratch. This design anchors subsequent generation to realistic acoustic patterns and discourse usage, reducing the risk that purely synthetic samples exhibit systematic artifacts or mismatched event distributions.
We choose \textit{InstructTTSEval}~\cite{instructttseval2025} as the seed source because it is an English--Chinese bilingual corpus with highly \emph{expressive} recordings and accompanying free-form captions. Importantly, the dataset provides free-form captions that may not explicitly describe NVVs and does not include predefined NVV tag annotations, allowing us to flexibly identify and label a broader range of NVV types rather than being limited to a fixed inventory.

We use Gemini 2.5 Pro as a multimodal annotator to jointly perform (i) NVV identification from audio, (ii) span-level localization by inserting inline markers into the transcript, and (iii) caption rewriting that explicitly mentions the NVVs if NVVs exist. This choice is motivated by the limitations of existing NVV recognition models, which typically predict only the event type without reliable text alignment and are trained on relatively restricted NVV taxonomies, making them less suitable for the broad NVV coverage.
In preliminary trials, we found that  Gemini 2.5 Pro is not sufficiently reliable: it may hallucinate subtle or nonexistent NVVs, or over-label events that are not perceptually salient. Therefore, to obtain high-confidence seeds, three annotators independently audit each candidate and judge whether the labeled NVV was clearly perceivable; only samples confirmed by the majority were retained, with disagreements adjudicated by a fourth reviewer. This process yields approximately 80 English and 30 Chinese seed instances. Although limited in scale, these seeds calibrate our generation prompts and serve as exemplars for subsequent stages.

\vspace{-1.5pt}
\textbf{Stage II: Taxonomy-driven controlled generation.}
For each NVV type in the 45-type taxonomy, we prompt Gemini 2.5 Pro to generate \emph{text-only} candidates in English and Chinese following a unified four-field schema: \texttt{text}, \texttt{text\_with\_nvv}, \texttt{caption\_with\_nvv}, and \texttt{nvv\_list}.
Generation is conditioned on a \emph{single target type} and we enforce a \emph{single-type constraint} by requiring \texttt{nvv\_list} to contain exactly one label that matches the target NVV type, while allowing the target NVV to occur multiple times within the same sentence when contextually natural.
Prompts encourage naturalistic and clearly perceivable contexts, discourage placeholders (e.g., ``[sound]''), and instruct captions to describe the NVV in natural language rather than tag tokens.
To promote diversity, we vary discourse settings (e.g., dialogue, narration, and instruction) and stylistic cues (e.g., neutral, excited, and calm) across batches.
To reduce redundancy, we perform near-duplicate removal via case-insensitive exact matching on \texttt{text\_with\_nvv}.

\vspace{-1.5pt}
\textbf{Stage III: Validation and replenishment.}
All generated candidates undergo an iterative two-stage validation loop.
First, automatic consistency checks verify schema conformity, ensure the declared NVV label matches the target type, filter out ambiguous or mixed-type cases, and validate marker consistency (e.g., \texttt{text\_with\_nvv} contains required inline tags and can be normalized back to \texttt{text}).
Second, annotators perform manual quality control to assess cross-field consistency among \texttt{text}, \texttt{text\_with\_nvv}, and \texttt{caption\_with\_nvv}, and remove samples that are contextually implausible or suggest weak perceptibility (e.g., speculative descriptions such as ``he might cough'' instead of an explicit cough event).
For NVVs that are easily confounded with textual interjections, we additionally screen the raw \texttt{text} and remove samples containing explicit interjection tokens (e.g., ``ah'', ``oh'', ``uh'', ``um'', ``hmm''). For NVV types with fewer than 50 validated instances after filtering, we trigger supplementary generation conditioned on that type and repeat validation until the target per-class quota is met. 

\textbf{Final dataset.}
After validation and iterative replenishment, we obtain $45 \times 50 = 2{,}250$ validated instances per language, yielding 2,250 English and 2,250 Chinese items. The resulting corpus contains 4,500 high-quality NVV instances with balanced class coverage and equal per-type representation. This curated resource provides a controlled foundation for developing and evaluating NVV-aware speech generation systems, as well as for studying NVV recognition and analysis.

\vspace{-7pt}
\subsection{Evaluation Protocol}
\vspace{-4pt}
\subsubsection{Objective metrics}
\label{sec:objective-metrics}
\vspace{-4pt}
For both prompt-based and tag-based systems, we report intelligibility using word error rate (WER)/character error rate (CER), and speech quality using DNSMOS P.835~\cite{Reddy2021DNSMOSP835}.
These metrics characterize the generated speech at the linguistic and signal levels, independent of explicit NVV controllability. Additionally, we compute the CLAP score~\cite{Elizalde2023CLAP} to measure caption-speech alignment for prompt-based systems, and NVV precision, recall, F1, and normalized tag distance to evaluate tag-based NVV controllability. All objective metrics are calculated from samples synthesized three times to assess stability. In contrast, subjective listening and LLM-as-a-judge evaluations are conducted on one run, due to the significantly higher cost of human and LLM assessments.

\vspace{0.2em}
\noindent\textbf{WER/CER$\downarrow$.}
We compute WER for English and CER for Chinese via automatic speech recognition (ASR). Specifically, English utterances are transcribed with \texttt{Whisper-large-v3}~\cite{radford23_whisper}, whereas Chinese utterances are transcribed with \texttt{paraformer-zh}~\cite{gao22_paraformer}.


\vspace{0.2em}
\noindent\textbf{DNSMOS P.835 (OVRL / SIG / BAK)$\uparrow$.}
DNSMOS P.835 is a non-intrusive perceptual metric that predicts ITU-T P.835-style components: SIG (speech signal quality), BAK (background noise intrusiveness), and OVRL (overall quality). 

\vspace{0.35em}
\noindent\textbf{Caption-speech alignment$\uparrow$ (prompt-based only).}
To evaluate caption-speech semantic alignment for prompt-based systems, we compute CLAP Score, defined as the cosine similarity between frozen CLAP embeddings of the synthesized speech and the corresponding caption text. 

\vspace{0.35em}
\noindent\textbf{NVV controllability (tag-based only).}
For tag-controlled synthesis, we quantify NVV controllability along two complementary axes: \emph{type correctness} and \emph{placement accuracy}.
Each NVV is represented as a tuple $(t,s)$, where $t\in\mathcal{T}$ denotes the NVV type and $s$ denotes its insertion position in the transcript.


We propose a \emph{GT-conditioned} verification method to infer NVV occurrences in synthesized speech using Gemini~2.5 Pro. In preliminary experiments, we found that directly prompting Gemini to detect NVVs from speech in an open-ended manner is unreliable, mainly due to hallucinated events and confusion among acoustically similar categories. 
For each utterance $u$, the Gemini verifier is provided with (i) the synthesized speech, (ii) the \emph{unchanged} reference transcript, and (iii) the \emph{target} NVV type $t_u$ specified by the ground-truth tag.
The verifier then outputs a binary presence decision $\texttt{present}_u\in\{0,1\}$ for the target NVV.
If $\texttt{present}_u=1$, it inserts \emph{exactly one} inline marker \texttt{<}$t_u$\texttt{>} into the reference transcript without any paraphrasing; otherwise, it returns the original transcript without any tag.
This constrained editing yields a deterministic predicted position index $s_{p,u}$ from the tagged transcript, producing a predicted tuple $(t_u,s_{p,u})$.
The verifier may additionally report a small set of \emph{non-target} NVV types; these are treated as spurious predictions when computing false positives.

\vspace{0.15em}
Let $(t_{g,u}, s_{g,u})$ denote the ground-truth NVV tuple for utterance $u$ (with type $t_{g,u}$ and position index $s_{g,u}$), and let $(t_{p,u}, s_{p,u})$ denote the predicted tuple returned by the verifier.
A predicted NVV is considered a match if and only if
\vspace{-5pt}
\[
\textstyle
t_{p,u}=t_{g,u}\quad\text{and}\quad |s_{p,u}-s_{g,u}|\le \delta,
\]
where $\delta$ is a fixed position tolerance measured in transcript indices~\footnote{Words for English and characters for Chinese.}.

\vspace{0.2em}
\noindent\textbf{Coverage $\uparrow$.}
We define coverage as the proportion of NVV instances supported by each speech generation system. For each system, we report the number of supported NVV types, denoted by $N_{\text{NVV\_supported}}$.
The coverage is then computed as:
\vspace{-6pt}
\[
\textstyle
\text{Cov.} = \frac{N_{\text{NVV\_supported}} \times 50}{45 \times 50}.
\]
This reflects the fraction of the total 2250 instances (45 NVV types × 50 instances per type) supported by the system.

\vspace{0.2em}
\noindent\textbf{Precision / Recall / F1$\uparrow$.}
Let $\mathrm{TP}$ be the number of utterances whose predicted NVV matches the ground-truth NVV under the above rule.
Let $\mathrm{FP}$ be the number of predicted NVVs that do not match any ground-truth NVV, including (i) target-type predictions with $|s_{p,u}-s_{g,u}|>\delta$ and (ii) any spurious non-target NVV predictions reported by the verifier.
Let $\mathrm{FN}$ be the number of ground-truth NVVs that are not matched by any prediction (e.g., $\texttt{present}_u=0$ or mismatched position beyond $\delta$).
We compute
\vspace{-5pt}
{\small
\[
\textstyle
\mathrm{Prec.}=\tfrac{\mathrm{TP}}{\mathrm{TP}+\mathrm{FP}},\quad
\mathrm{Rec.}=\tfrac{\mathrm{TP}}{\mathrm{TP}+\mathrm{FN}},\quad
\mathrm{F1}=\tfrac{2\,\mathrm{TP}}{2\,\mathrm{TP}+\mathrm{FP}+\mathrm{FN}}.
\]
}
These scores capture NVV type correctness jointly with placement accuracy under tolerance $\delta$.

\vspace{0.2em}
\noindent\textbf{Normalized tag distance (NTD)$\downarrow$.}
For each utterance $u$ whose NVV is matched (i.e., contributes to $\mathrm{TP}$), define the absolute position error
$d_u = |s_{p,u}-s_{g,u}|$ and let $L_u$ denote the transcript length.
We report the length-normalized mean position error over matched utterances:
\vspace{-5pt}
\[
\textstyle
\mathrm{NTD}
=\frac{1}{|\mathcal{U}_{\mathrm{TP}}|}\sum_{u\in \mathcal{U}_{\mathrm{TP}}}\frac{d_u}{L_u},
\]
where $\mathcal{U}_{\mathrm{TP}}=\{u:\ (t_{p,u},s_{p,u})\ \text{matches}\ (t_{g,u},s_{g,u})\}$ denotes the set of matched utterances.
Lower NTD indicates a more accurate placement relative to the utterance length.

\vspace{-5pt}
\subsubsection{Subjective metrics}
\label{sec:subjective-metrics}
\vspace{-4pt}
We conduct human listening tests on 450 randomly selected samples (10 per NVV type) per language via the Prolific platform~\footnote{\url{https://app.prolific.com/}}. We assess naturalness, quality, and NVV Perceptual Effect (NVV PE) for both prompt-based and tag-based systems. In addition, we evaluate overall instruction following (IF), NVV Instruction Following (NVV IF) for prompt-based systems, and NVV Accuracy and expression for tag-based systems. All subjective criteria are rated on a 5-point Likert scale (5 = best, 1 = worst). To explicitly capture complete NVV failures, we include a score of 0 for NVV-specific criteria—such as NVV IF, NVV Accuracy, and NVV PE—when the target NVV is absent or nearly inaudible.

\begin{table*}[ht]
\centering
\vspace{-5pt}
\caption{Objective results of prompt-based and tag-based systems across three independent generation runs. 
Best is in bold and second-best is underlined within each block, excluding the ``(w/o NVV)'' rows. ``--'' indicates not applicable.
}

\label{tab:prompt_tag_obj_res}
\scriptsize
\setlength{\tabcolsep}{1.2pt}
\renewcommand{\arraystretch}{0.98}
\setlength{\extrarowheight}{1pt}  
\resizebox{0.8\textwidth}{!}{%
 \scalebox{0.6}{ 
\begin{tabular}{l c c c c c c c c c c c}
\toprule
\multirow{2}{*}{\textbf{System}} &
\multirow{2}{*}{\textbf{Lang}} &
\multirow{2}{*}{\textbf{WER/CER}$\downarrow$} &
\multicolumn{3}{c}{\textbf{DNSMOS}$\uparrow$} &
\multirow{2}{*}{\textbf{CLAP Score}$\uparrow$} &
\multirow{2}{*}{\textbf{Coverage}$\uparrow$} &
\multirow{2}{*}{\textbf{Precision}$\uparrow$} &
\multirow{2}{*}{\textbf{Recall}$\uparrow$} &
\multirow{2}{*}{\textbf{F1}$\uparrow$} &
\multirow{2}{*}{\textbf{NTD}$\downarrow$} \\
\cmidrule(lr){4-6}
& & & \textbf{SIG} & \textbf{BAK} & \textbf{OVRL} & & & & & & \\[0.4ex]
\midrule

\rowcolor{cyan!8}\multicolumn{12}{c}{\textbf{Prompt-based Systems}}\\[-0.2ex]

Parler-TTS Mini
& EN & 6.25 {\scriptsize $\pm$ 0.36}
& 3.48 {\scriptsize $\pm$ 0.00} & 4.03 {\scriptsize $\pm$ 0.00} & 3.20 {\scriptsize $\pm$ 0.00}
& 0.34 {\scriptsize $\pm$ 0.00} & -- & -- & -- & -- & -- \\

Parler-TTS Large
& EN & 9.30 {\scriptsize $\pm$ 1.40}
& 3.45 {\scriptsize $\pm$ 0.00} & 4.00 {\scriptsize $\pm$ 0.00} & 3.15 {\scriptsize $\pm$ 0.00}
& 0.35 {\scriptsize $\pm$ 0.00} & -- & -- & -- & -- & -- \\

CapSpeech
& EN & \underline{4.56 {\scriptsize $\pm$ 0.04}}
& 3.46 {\scriptsize $\pm$ 0.00} & 3.96 {\scriptsize $\pm$ 0.00} & 3.16 {\scriptsize $\pm$ 0.00}
& 0.40 {\scriptsize $\pm$ 0.00} & -- & -- & -- & -- & -- \\

Qwen3-TTS
& EN & \textbf{2.06 {\scriptsize $\pm$ 0.03}}
& 3.56 {\scriptsize $\pm$ 0.00} & \underline{4.07 {\scriptsize $\pm$ 0.00}} & \underline{3.30 {\scriptsize $\pm$ 0.00}}
& \textbf{0.45 {\scriptsize $\pm$ 0.00}} & -- & -- & -- & -- & -- \\

GPT-4o mini TTS
& EN & 4.81 {\scriptsize $\pm$ 0.67}
& \textbf{3.59 {\scriptsize $\pm$ 0.00}} & \textbf{4.14 {\scriptsize $\pm$ 0.00}} & \textbf{3.35 {\scriptsize $\pm$ 0.00}}
& \underline{0.44 {\scriptsize $\pm$ 0.00}} & -- & -- & -- & -- & -- \\

Gemini 2.5 Flash
& EN & 58.80 {\scriptsize $\pm$ 19.52}
& \underline{3.57 {\scriptsize $\pm$ 0.00}} & 4.02 {\scriptsize $\pm$ 0.00} & 3.29 {\scriptsize $\pm$ 0.03}
& 0.42 {\scriptsize $\pm$ 0.00} & -- & -- & -- & -- & -- \\

Gemini 2.5 Pro
& EN & 5.40 {\scriptsize $\pm$ 0.28}
& 3.52 {\scriptsize $\pm$ 0.00} & 4.00 {\scriptsize $\pm$ 0.00} & 3.23 {\scriptsize $\pm$ 0.00}
& 0.41 {\scriptsize $\pm$ 0.00} & -- & -- & -- & -- & -- \\

\g Gemini 2.5 Pro {\scriptsize (w/o NVV)}
& \g EN & \g 3.65 {\scriptsize $\pm$ 0.00}
& \g 3.52 {\scriptsize $\pm$ 0.00} & \g 4.01 {\scriptsize $\pm$ 0.00} & \g 3.24 {\scriptsize $\pm$ 0.00}
& \g 0.42 {\scriptsize $\pm$ 0.00}  & \g -- & \g -- & \g -- & \g -- & \g -- \\

\cmidrule(lr){1-12}

Qwen3-TTS
& ZH & \textbf{4.08 {\scriptsize $\pm$ 0.07}}
& 3.50 {\scriptsize $\pm$ 0.00} & 3.98 {\scriptsize $\pm$ 0.00} & 3.19 {\scriptsize $\pm$ 0.00}
& 0.39 {\scriptsize $\pm$ 0.00} & -- & -- & -- & -- & -- \\

GPT-4o mini TTS
& ZH & \underline{4.67 {\scriptsize $\pm$ 0.23}}
& \textbf{3.64 {\scriptsize $\pm$ 0.00}} & \textbf{4.16 {\scriptsize $\pm$ 0.00}} & \textbf{3.40 {\scriptsize $\pm$ 0.00}}
& \textbf{0.43 {\scriptsize $\pm$ 0.00}} & -- & -- & -- & -- & -- \\

Gemini 2.5 Flash
& ZH & 16.45 {\scriptsize $\pm$ 1.76}
& \underline{3.57 {\scriptsize $\pm$ 0.00}} & 3.97 {\scriptsize $\pm$ 0.00} & 3.25 {\scriptsize $\pm$ 0.00}
& 0.41 {\scriptsize $\pm$ 0.00} & -- & -- & -- & -- & -- \\

Gemini 2.5 Pro
& ZH & 7.68 {\scriptsize $\pm$ 0.22}
& 3.55 {\scriptsize $\pm$ 0.00} & \underline{4.01 {\scriptsize $\pm$ 0.00}} & \underline{3.26 {\scriptsize $\pm$ 0.00}}
& \underline{0.42 {\scriptsize $\pm$ 0.00}} & -- & -- & -- & -- & -- \\

\g Gemini 2.5 Pro {\scriptsize (w/o NVV)}
& \g ZH & \g 5.88 {\scriptsize $\pm$ 0.00}
& \g 3.52 {\scriptsize $\pm$ 0.00} & \g 4.01 {\scriptsize $\pm$ 0.00} & \g 3.24 {\scriptsize $\pm$ 0.00}
& \g 0.43 {\scriptsize $\pm$ 0.00}  & \g -- & \g -- & \g -- & \g -- & \g -- \\

\midrule

\rowcolor{cyan!8}\multicolumn{12}{c}{\textbf{Tag-based Systems}}\\[-0.2ex]

Bark
& EN & 14.73 {\scriptsize $\pm$ 1.53}
& 3.00 {\scriptsize $\pm$ 0.03} & 3.08 {\scriptsize $\pm$ 0.03} & 2.52 {\scriptsize $\pm$ 0.00}
& -- & 0.11
& 0.614 {\scriptsize $\pm$ 0.040}
& 0.700 {\scriptsize $\pm$ 0.043}
& 0.654 {\scriptsize $\pm$ 0.041}
& 0.0037 {\scriptsize $\pm$ 0.0000} \\

Higgs-Audio
& EN & 9.41 {\scriptsize $\pm$ 3.77}
& 3.55 {\scriptsize $\pm$ 0.00} & 3.96 {\scriptsize $\pm$ 0.03} & 3.23 {\scriptsize $\pm$ 0.00}
& -- & 0.09
& 0.360 {\scriptsize $\pm$ 0.053}
& 0.407 {\scriptsize $\pm$ 0.047}
& 0.382 {\scriptsize $\pm$ 0.050}
& 0.0111 {\scriptsize $\pm$ 0.0033} \\

ChatTTS
& EN & 5.58 {\scriptsize $\pm$ 0.90}
& \textbf{3.67 {\scriptsize $\pm$ 0.00}} & \underline{4.12 {\scriptsize $\pm$ 0.03}} & \underline{3.42 {\scriptsize $\pm$ 0.03}}
& -- & 0.02
& 0.652 {\scriptsize $\pm$ 0.140}
& 0.680 {\scriptsize $\pm$ 0.203}
& 0.664 {\scriptsize $\pm$ 0.167}
& \textbf{0.0028 {\scriptsize $\pm$ 0.0028}} \\

Fish-Speech
& EN & 5.65 {\scriptsize $\pm$ 1.36}
& 3.55 {\scriptsize $\pm$ 0.00} & 4.06 {\scriptsize $\pm$ 0.00} & 3.28 {\scriptsize $\pm$ 0.00}
& -- & 0.16
& 0.447 {\scriptsize $\pm$ 0.023}
& 0.418 {\scriptsize $\pm$ 0.024}
& 0.432 {\scriptsize $\pm$ 0.024}
& 0.0157 {\scriptsize $\pm$ 0.0022} \\

Dia
& EN & 21.95 {\scriptsize $\pm$ 1.14}
& 2.94 {\scriptsize $\pm$ 0.00} & 2.71 {\scriptsize $\pm$ 0.00} & 2.27 {\scriptsize $\pm$ 0.00}
& -- & \textbf{0.29}
& 0.574 {\scriptsize $\pm$ 0.010}
& 0.705 {\scriptsize $\pm$ 0.006}
& 0.632 {\scriptsize $\pm$ 0.008}
& 0.0052 {\scriptsize $\pm$ 0.0000} \\

CosyVoice 2
& EN & \underline{3.82 {\scriptsize $\pm$ 0.18}}
& \textbf{3.67 {\scriptsize $\pm$ 0.00}} & \textbf{4.19 {\scriptsize $\pm$ 0.00}} & \textbf{3.45 {\scriptsize $\pm$ 0.00}}
& -- & 0.18
& 0.475 {\scriptsize $\pm$ 0.024}
& 0.451 {\scriptsize $\pm$ 0.025}
& 0.463 {\scriptsize $\pm$ 0.024}
& 0.0159 {\scriptsize $\pm$ 0.0042} \\

Orpheus TTS
& EN & 4.98 {\scriptsize $\pm$ 0.34}
& \underline{3.62 {\scriptsize $\pm$ 0.00}} & 4.11 {\scriptsize $\pm$ 0.00} & 3.34 {\scriptsize $\pm$ 0.00}
& -- & 0.18
& \textbf{0.687 {\scriptsize $\pm$ 0.026}}
& \underline{0.774 {\scriptsize $\pm$ 0.034}}
& \textbf{0.728 {\scriptsize $\pm$ 0.029}}
& \underline{0.0031 {\scriptsize $\pm$ 0.0000}} \\

ElevenLabs
& EN & \textbf{2.31 {\scriptsize $\pm$ 0.59}}
& 3.59 {\scriptsize $\pm$ 0.11} & 4.06 {\scriptsize $\pm$ 0.08} & 3.32 {\scriptsize $\pm$ 0.14}
& -- & \underline{0.27}
& \underline{0.664 {\scriptsize $\pm$ 0.017}}
& \textbf{0.787 {\scriptsize $\pm$ 0.034}}
& \underline{0.720 {\scriptsize $\pm$ 0.024}}
& 0.0091 {\scriptsize $\pm$ 0.0014} \\

\g ElevenLabs {\scriptsize (w/o NVV)}
& \g EN & \g 2.16 {\scriptsize $\pm$ 0.00}
& \g 3.70 {\scriptsize $\pm$ 0.00} & \g 4.17 {\scriptsize $\pm$ 0.00} & \g 3.47 {\scriptsize $\pm$ 0.00}
& \g -- & \g -- & \g -- & \g -- & \g -- & \g -- \\

\cmidrule(lr){1-12}

Bark
& ZH & 41.86 {\scriptsize $\pm$ 2.14}
& 2.99 {\scriptsize $\pm$ 0.00} & 3.10 {\scriptsize $\pm$ 0.03} & 2.54 {\scriptsize $\pm$ 0.00}
& -- & 0.11
& 0.572 {\scriptsize $\pm$ 0.032}
& 0.605 {\scriptsize $\pm$ 0.016}
& 0.588 {\scriptsize $\pm$ 0.017}
& \textbf{0.0141 {\scriptsize $\pm$ 0.0069}} \\

Orpheus TTS
& ZH & 18.78 {\scriptsize $\pm$ 0.59}
& 3.60 {\scriptsize $\pm$ 0.00} & \underline{4.11 {\scriptsize $\pm$ 0.00}} & 3.30 {\scriptsize $\pm$ 0.00}
& -- & \underline{0.18}
& 0.585 {\scriptsize $\pm$ 0.020}
& 0.671 {\scriptsize $\pm$ 0.021}
& 0.625 {\scriptsize $\pm$ 0.020}
& \underline{0.0159 {\scriptsize $\pm$ 0.0030}} \\

Higgs-Audio
& ZH & 5.98 {\scriptsize $\pm$ 1.08}
& 3.52 {\scriptsize $\pm$ 0.00} & 3.91 {\scriptsize $\pm$ 0.03} & 3.17 {\scriptsize $\pm$ 0.03}
& -- & 0.09
& 0.429 {\scriptsize $\pm$ 0.013}
& 0.369 {\scriptsize $\pm$ 0.021}
& 0.396 {\scriptsize $\pm$ 0.016}
& 0.0186 {\scriptsize $\pm$ 0.0035} \\

Fish-Speech
& ZH & 9.25 {\scriptsize $\pm$ 0.45}
& 3.49 {\scriptsize $\pm$ 0.00} & 4.00 {\scriptsize $\pm$ 0.00} & 3.19 {\scriptsize $\pm$ 0.00}
& -- & 0.16
& 0.592 {\scriptsize $\pm$ 0.004}
& 0.605 {\scriptsize $\pm$ 0.011}
& 0.598 {\scriptsize $\pm$ 0.003}
& 0.0397 {\scriptsize $\pm$ 0.0037} \\

ChatTTS
& ZH & \underline{4.52 {\scriptsize $\pm$ 0.91}}
& 3.49 {\scriptsize $\pm$ 0.00} & 3.50 {\scriptsize $\pm$ 0.58} & 2.94 {\scriptsize $\pm$ 0.29}
& -- & 0.02
& \textbf{0.644 {\scriptsize $\pm$ 0.045}}
& \textbf{0.773 {\scriptsize $\pm$ 0.064}}
& \textbf{0.703 {\scriptsize $\pm$ 0.052}}
& 0.0237 {\scriptsize $\pm$ 0.0092} \\

CosyVoice 2
& ZH & 6.27 {\scriptsize $\pm$ 1.07}
& \textbf{3.63 {\scriptsize $\pm$ 0.00}} & \textbf{4.17 {\scriptsize $\pm$ 0.00}} & \textbf{3.40 {\scriptsize $\pm$ 0.00}}
& -- & \underline{0.18}
& 0.515 {\scriptsize $\pm$ 0.014}
& 0.480 {\scriptsize $\pm$ 0.018}
& 0.496 {\scriptsize $\pm$ 0.016}
& 0.0383 {\scriptsize $\pm$ 0.0024} \\

ElevenLabs
& ZH & \textbf{4.13 {\scriptsize $\pm$ 0.87}}
& \underline{3.61 {\scriptsize $\pm$ 0.04}} & 4.03 {\scriptsize $\pm$ 0.00} & \underline{3.32 {\scriptsize $\pm$ 0.04}}
& -- & \textbf{0.27}
& \underline{0.630 {\scriptsize $\pm$ 0.025}}
& \underline{0.750 {\scriptsize $\pm$ 0.041}}
& \underline{0.684 {\scriptsize $\pm$ 0.032}}
& 0.0246 {\scriptsize $\pm$ 0.0028} \\
\g ElevenLabs {\scriptsize (w/o NVV)}
& \g ZH & \g 2.38 {\scriptsize $\pm$ 0.00}
& \g 3.67 {\scriptsize $\pm$ 0.00} & \g 4.08 {\scriptsize $\pm$ 0.00} & \g 3.39 {\scriptsize $\pm$ 0.00}
& \g -- & \g -- & \g -- & \g -- & \g -- & \g -- \\

\bottomrule
\end{tabular}
}
}
\vspace{-5pt}
\end{table*}

\vspace{0.2em}
\noindent\textbf{Overall naturalness$\uparrow$.}
This criterion measures perceived human-likeness and prosodic fluency of the spoken content, including speaking rate, pausing, emphasis, and intonation patterns, while abstracting away from signal-level artifacts.

\vspace{0.2em}
\noindent\textbf{Overall quality$\uparrow$.}
This criterion measures signal-level fidelity independent of content and style, focusing on background noise, distortion/clipping, reverberation, loudness instability, and codec-related artifacts.

\vspace{0.2em}
\noindent\textbf{Overall instruction following (IF)$\uparrow$ (prompt-based only).}
IF measures global consistency between the synthesized speech and the caption in speaker attributes and speaking style (e.g., timbre, age, gender, emotional tone, and scene atmosphere). 

\vspace{0.2em}
\noindent\textbf{Overall expression$\uparrow$ (tag-based only).}
This criterion measures the overall effectiveness of the utterance as a coherent performance, considering speech prosody and realized NVVs jointly, without explicitly referencing the caption text.

\vspace{0.2em}
\noindent\textbf{NVV instruction following (IF)$\uparrow$ (prompt-based only).}
NVV IF measures compliance with NVV-related instructions specified in the caption, including whether the intended NVV type is realized and whether it occurs in an approximately appropriate region, while penalizing omissions and salient spurious NVVs. A score of 0 indicates that the target NVV is absent or almost inaudible.

\vspace{0.2em}
\noindent\textbf{NVV accuracy$\uparrow$ (tag-based only).}
NVV Accuracy measures adherence to the input tag in terms of NVV type and coarse placement, while penalizing omissions and salient spurious NVVs. A score of 0 indicates that the tagged NVV is absent or almost inaudible.

\vspace{0.25em}
\noindent\textbf{NVV perceptual effect (PE)$\uparrow$.}
NVV PE measures the perceived naturalness and expressive effectiveness of the NVV in the generated speech, including whether it sounds human-like and integrates smoothly with surrounding speech. A score of 0 indicates that the target NVV is absent or almost inaudible.

\begin{table*}[!htbp]
\vspace{-10pt}
\caption{Subjective results of prompt-based and tag-based systems.
Scores are reported as mean $\pm$ 95\% confidence interval.
``--'' indicates not applicable. Best is in \textbf{bold} and second-best is \underline{underlined} \emph{within each language and each system type}.}
\centering
\label{tab:merged_sub_res}
\scriptsize
\setlength{\tabcolsep}{4pt}
\renewcommand{\arraystretch}{1.08}

\resizebox{0.8\textwidth}{!}{%
\begin{tabular}{llccccccc}
\toprule
\textbf{System} & \textbf{Lang} &
\shortstack[c]{\textbf{Overall Naturalness}$\uparrow$} &
\shortstack[c]{\textbf{Overall Quality}$\uparrow$} &
\shortstack[c]{\textbf{NVV PE}$\uparrow$} &
\shortstack[c]{\textbf{Overall IF}$\uparrow$} &
\shortstack[c]{\textbf{NVV IF}$\uparrow$} &
\shortstack[c]{\textbf{NVV Accuracy}$\uparrow$} &
\shortstack[c]{\textbf{Overall Expression}$\uparrow$} \tabularnewline
\midrule

\rowcolor{cyan!8}\multicolumn{9}{c}{\textbf{Prompt-based Systems}} \tabularnewline[-0.2ex]

Parler-TTS Large & EN & 2.69$\pm$0.08 & 3.35$\pm$0.09 & 0.93$\pm$0.09 & 2.52$\pm$0.08 & 0.99$\pm$0.09 & -- & -- \tabularnewline
Parler-TTS Mini  & EN & 2.62$\pm$0.08 & 3.40$\pm$0.09 & 0.85$\pm$0.09 & 2.44$\pm$0.08 & 0.92$\pm$0.09 & -- & -- \tabularnewline
CapSpeech        & EN & 2.73$\pm$0.08 & 3.41$\pm$0.09 & 1.01$\pm$0.09 & 2.63$\pm$0.08 & 1.11$\pm$0.10 & -- & -- \tabularnewline
GPT-4o mini TTS  & EN & 3.33$\pm$0.08 & 3.56$\pm$0.08 & 1.74$\pm$0.12 & 3.20$\pm$0.09 & 1.89$\pm$0.12 & -- & -- \tabularnewline
Qwen3-TTS        & EN & 3.44$\pm$0.09 & 3.86$\pm$0.08 & 2.03$\pm$0.14 & 3.58$\pm$0.09 & 2.15$\pm$0.14 & -- & -- \tabularnewline
Gemini 2.5 Flash & EN & \underline{4.00$\pm$0.07} & \underline{4.28$\pm$0.07} & \underline{2.60$\pm$0.13} & \underline{3.86$\pm$0.08} & \underline{2.67$\pm$0.13} & -- & -- \tabularnewline
Gemini 2.5 Pro   & EN & \textbf{4.07$\pm$0.07} & \textbf{4.30$\pm$0.06} & \textbf{2.68$\pm$0.13} & \textbf{3.92$\pm$0.07} & \textbf{2.74$\pm$0.13} & -- & -- \tabularnewline

\cmidrule(lr){1-9}

GPT-4o mini TTS  & ZH & 2.11$\pm$0.09 & 2.77$\pm$0.12 & 0.84$\pm$0.11 & 2.46$\pm$0.12 & 0.92$\pm$0.12 & -- & -- \tabularnewline
Qwen3-TTS        & ZH & \textbf{3.45$\pm$0.11} & \textbf{3.93$\pm$0.10} & 2.01$\pm$0.16 & \textbf{3.40$\pm$0.11} & 1.98$\pm$0.16 & -- & -- \tabularnewline
Gemini 2.5 Flash & ZH & 3.04$\pm$0.11 & 3.71$\pm$0.10 & \textbf{2.29$\pm$0.14} & 3.16$\pm$0.11 & \textbf{2.42$\pm$0.15} & -- & -- \tabularnewline
Gemini 2.5 Pro   & ZH & \underline{3.38$\pm$0.10} & \underline{3.75$\pm$0.10} & \underline{2.07$\pm$0.15} & \underline{3.32$\pm$0.11} & \underline{2.11$\pm$0.16} & -- & -- \tabularnewline

\midrule

\rowcolor{cyan!8}\multicolumn{9}{c}{\textbf{Tag-based Systems}} \tabularnewline[-0.2ex]

Bark         & EN & 2.75$\pm$0.39 & 2.37$\pm$0.34 & 2.07$\pm$0.49 & -- & -- & 2.65$\pm$0.47 & 2.62$\pm$0.41 \tabularnewline
Dia          & EN & 3.12$\pm$0.20 & 2.82$\pm$0.19 & 2.45$\pm$0.25 & -- & -- & 2.99$\pm$0.25 & 3.24$\pm$0.20 \tabularnewline
Fish-Speech  & EN & 3.53$\pm$0.26 & 3.62$\pm$0.28 & 1.01$\pm$0.30 & -- & -- & 1.04$\pm$0.30 & 2.81$\pm$0.26 \tabularnewline
Higgs-Audio  & EN & 3.63$\pm$0.32 & 3.63$\pm$0.36 & 2.41$\pm$0.45 & -- & -- & 2.28$\pm$0.49 & 3.28$\pm$0.32 \tabularnewline
CosyVoice 2  & EN & 3.65$\pm$0.22 & 3.92$\pm$0.24 & 2.39$\pm$0.35 & -- & -- & 2.22$\pm$0.36 & 3.34$\pm$0.26 \tabularnewline
ChatTTS      & EN & 3.30$\pm$0.75 & 3.10$\pm$0.79 & 3.00$\pm$0.73 & -- & -- & 3.40$\pm$0.73 & 2.95$\pm$0.80 \tabularnewline
Orpheus TTS  & EN & \underline{4.01$\pm$0.28} & \underline{4.11$\pm$0.27} & \underline{3.31$\pm$0.32} & -- & -- & \underline{3.71$\pm$0.32} & \underline{3.49$\pm$0.29} \tabularnewline
ElevenLabs   & EN & \textbf{4.60$\pm$0.10} & \textbf{4.71$\pm$0.10} & \textbf{3.92$\pm$0.22} & -- & -- & \textbf{4.21$\pm$0.20} & \textbf{4.28$\pm$0.14} \tabularnewline

\cmidrule(lr){1-9}

Bark         & ZH & 2.23$\pm$0.22 & 2.00$\pm$0.20 & 1.09$\pm$0.25 & -- & -- & 1.12$\pm$0.28 & 2.02$\pm$0.21 \tabularnewline
Fish-Speech  & ZH & 2.71$\pm$0.18 & 3.09$\pm$0.20 & 0.10$\pm$0.05 & -- & -- & 0.13$\pm$0.08 & 2.43$\pm$0.19 \tabularnewline
Orpheus TTS  & ZH & 3.20$\pm$0.17 & 3.29$\pm$0.16 & 2.05$\pm$0.28 & -- & -- & 2.17$\pm$0.28 & 2.92$\pm$0.17 \tabularnewline
Higgs-Audio  & ZH & 3.48$\pm$0.24 & 3.68$\pm$0.25 & 1.38$\pm$0.40 & -- & -- & 1.18$\pm$0.40 & 3.36$\pm$0.24 \tabularnewline
CosyVoice 2  & ZH & \underline{3.76$\pm$0.14} & \textbf{4.35$\pm$0.13} & 1.56$\pm$0.26 & -- & -- & 1.65$\pm$0.29 & 3.28$\pm$0.16 \tabularnewline
ChatTTS      & ZH & 3.53$\pm$0.36 & 3.13$\pm$0.36 & \underline{2.13$\pm$0.59} & -- & -- & \underline{2.23$\pm$0.58} & \underline{3.40$\pm$0.40} \tabularnewline
ElevenLabs   & ZH & \textbf{4.09$\pm$0.11} & \underline{4.31$\pm$0.10} & \textbf{3.38$\pm$0.25} & -- & -- & \textbf{3.41$\pm$0.26} & \textbf{3.98$\pm$0.12} \tabularnewline

\bottomrule
\end{tabular}%
}
\end{table*}

\vspace{-4pt}

\vspace{-5pt}
\subsubsection{LLM-based multi-rater}
\label{sec:llm-judge}
\vspace{-4pt}
Recent advancements in audio-aware large language models (LLMs) have shown their potential as reliable evaluators in speech and audio assessments~\cite{wang2025speechllm,chiang2025audio}. In this work, we adopt the LLM-based multi-rater evaluation method to complement subjective listening tests. To ensure comparability, the same metrics and rating scales used in the subjective evaluation are applied.

To mitigate systematic biases and reduce score variance, we implement several key controls:
(i)  Anonymization and Randomization: candidate speech samples are anonymized (A/B/C) and their order is randomized to prevent bias from system identifiers.
(ii)  Strict Rubric Compliance: rubric compliance is enforced through an artifact inventory and score-capping constraints to avoid overly optimistic ratings due to common synthesis artifacts such as metallic timbre, clipping, and discontinuities.
(iii)  Reproducibility and Stability: a low sampling temperature (0.2) and a fixed random seed are used to ensure stable and reproducible evaluations. Additionally, evaluations are conducted in three rounds with a three-fold partition, ensuring fold-wise stability.
(iv) Multi-Rater Setup: each item is evaluated by a subset of four independent LLM raters, ensuring each sample is evaluated by at least one rater. Scores are aggregated across raters to reduce individual biases and improve evaluation reliability.
(v) Comparative Evaluation Mode: multiple systems are evaluated for the same task (tag or sample) with anonymized labels (A/B/C) and relative scores, which helps stabilize rankings and simulate human judgment more accurately.

By combining these controls, the LLM-based multi-rater evaluation method provides a systematic, reproducible, and scalable approach for evaluating speech and audio, complementing human evaluations while minimizing subjective bias and ensuring reliable results.

\vspace{-8pt}
\section{Speech Generation Systems}
\label{sec:eval-setup}
\vspace{-4pt}
Existing systems that support speech generation with NVVs can be divided into two categories: prompt-based and tag-based systems. We benchmark 15 speech generation systems, including 7 prompt-based and 8 tag-based systems, offering a diverse evaluation spectrum. Commercial models demonstrate strong industrial performance, while open-source systems emphasize transparency, reproducibility, and research into controllable speech generation.

\vspace{0.2em}
\textbf{Prompt-based speech generation}.
Prompt-based speech generation systems control speech generation through descriptive captions (\texttt{caption\_with\_nvv} field) that specify desired attributes, such as gender, age, emotion, speaking style, and NVVs. We benchmarked 7 prompt-based systems capable of NVV synthesis with 3 commercial and 4 open-source systems. The commercial systems include \textit{Gemini 2.5 Pro} and \textit{Gemini 2.5 Flash} by Google, and \textit{GPT-4o mini TTS} by OpenAI, all of which generate speech based on captions with pre-defined voices. On the open-source side, we considered \textit{Parler-TTS Mini}~\footnote{\href{https://huggingface.co/parler-tts/parler-tts-mini-v1}{https://huggingface.co/parler-tts/parler-tts-mini-v1}}, \textit{Parler-TTS Large}~\footnote{\href{https://huggingface.co/parler-tts/parler-tts-large-v1}{https://huggingface.co/parler-tts/parler-tts-large-v1}}, \textit{Qwen3-TTS}~\cite{hu2026qwen3}, and \textit{CapSpeech}~\cite{wang2025capspeech}.

\textbf{Tag-based speech generation}.
Tag-based systems generate speech with NVVs by inserting the corresponding NVV tags into the text (\texttt{text\_with\_nvv} field). We evaluated 8 tag-based systems listed in Table~\ref{tab:nvv_tags_and_datasets}, including 7 open-source systems \textit{ChatTTS}~\cite{chattts_github_2024}, \textit{Higgs-Audio}~\cite{higgsaudio_github_2025}, \textit{Bark}~\cite{bark_github_2023}, \textit{Fish-Speech}~\cite{liao2024fishspeech}, \textit{Orpheus TTS}~\cite{orpheus_tts_github_2025}, \textit{CosyVoice 2}~\cite{du2024cosyvoice2}, \textit{Dia}~\cite{dia_github}, and one commercial system \textit{ElevenLabs}~\cite{elevenlabs_models_docs}.

\begin{table*}[t]
\centering

\caption{LLM evaluation results of prompt-based and tag-based systems. Scores are reported as mean $\pm$ 95\% confidence interval. ``--'' indicates not applicable. Best is in \textbf{bold} and second-best is \underline{underlined} within each language and each system type.}
\label{tab:merged_llm_res}
\scriptsize
\setlength{\tabcolsep}{3pt}
\renewcommand{\arraystretch}{1.05}

\resizebox{0.8\textwidth}{!}{%
\begin{tabular}{llccccccc}
\toprule
\textbf{System} &
\textbf{Lang} &
\makecell{\textbf{Overall Naturalness}$\uparrow$} &
\makecell{\textbf{Overall Quality}$\uparrow$} &
\makecell{\textbf{NVV PE}$\uparrow$} &
\makecell{\textbf{Overall IF}$\uparrow$} &
\makecell{\textbf{NVV IF}$\uparrow$} &
\makecell{\textbf{NVV Accuracy}$\uparrow$} &
\makecell{\textbf{Overall Expression}$\uparrow$} \tabularnewline
\midrule

\rowcolor{cyan!8}\multicolumn{9}{c}{\textbf{Prompt-based Systems}} \tabularnewline

Parler-TTS Mini  & EN & 1.65$\pm$0.05 & 3.30$\pm$0.08 & 0.23$\pm$0.05 & 1.54$\pm$0.05 & 0.30$\pm$0.06 & -- & -- \tabularnewline
Parler-TTS Large & EN & 1.67$\pm$0.05 & 3.13$\pm$0.09 & 0.30$\pm$0.06 & 1.59$\pm$0.06 & 0.38$\pm$0.07 & -- & -- \tabularnewline
CapSpeech        & EN & 1.88$\pm$0.06 & 3.52$\pm$0.07 & 0.48$\pm$0.07 & 1.88$\pm$0.07 & 0.57$\pm$0.09 & -- & -- \tabularnewline
GPT-4o mini TTS  & EN & 3.16$\pm$0.07 & \underline{3.93$\pm$0.06} & 1.73$\pm$0.14 & 3.17$\pm$0.09 & 1.65$\pm$0.13 & -- & -- \tabularnewline
Qwen3-TTS        & EN & 3.17$\pm$0.07 & \textbf{3.96$\pm$0.06} & 2.16$\pm$0.15 & 3.44$\pm$0.09 & 1.98$\pm$0.14 & -- & -- \tabularnewline
Gemini 2.5 Flash & EN & \textbf{3.30$\pm$0.09} & 3.80$\pm$0.06 & \textbf{2.84$\pm$0.14} & \textbf{3.67$\pm$0.10} & \textbf{2.78$\pm$0.13} & -- & -- \tabularnewline
Gemini 2.5 Pro   & EN & \underline{3.20$\pm$0.08} & 3.65$\pm$0.06 & \underline{2.54$\pm$0.14} & \underline{3.53$\pm$0.09} & \underline{2.64$\pm$0.14} & -- & -- \tabularnewline

\cmidrule(lr){1-9}

GPT-4o mini TTS  & ZH & 2.50$\pm$0.06 & 3.85$\pm$0.06 & 0.92$\pm$0.10 & 2.38$\pm$0.08 & 1.04$\pm$0.11 & -- & -- \tabularnewline
Qwen3-TTS        & ZH & \underline{3.23$\pm$0.07} & \underline{3.88$\pm$0.06} & 2.10$\pm$0.14 & \underline{3.46$\pm$0.08} & 1.99$\pm$0.13 & -- & -- \tabularnewline
Gemini 2.5 Flash & ZH & \textbf{3.27$\pm$0.08} & \textbf{3.90$\pm$0.06} & \textbf{3.06$\pm$0.12} & \textbf{3.78$\pm$0.09} & \textbf{3.12$\pm$0.12} & -- & -- \tabularnewline
Gemini 2.5 Pro   & ZH & 3.11$\pm$0.07 & 3.73$\pm$0.06 & \underline{2.54$\pm$0.13} & 3.44$\pm$0.08 & \underline{2.71$\pm$0.13} & -- & -- \tabularnewline

\midrule

\rowcolor{cyan!8}\multicolumn{9}{c}{\textbf{Tag-based Systems}} \tabularnewline

Bark        & EN & 1.78$\pm$0.19 & 2.20$\pm$0.22 & 1.43$\pm$0.28 & -- & -- & 2.34$\pm$0.40 & 1.71$\pm$0.20 \tabularnewline
Dia         & EN & 2.19$\pm$0.16 & 2.51$\pm$0.19 & 2.08$\pm$0.24 & -- & -- & 3.14$\pm$0.30 & 2.49$\pm$0.21 \tabularnewline
Fish-Speech & EN & 1.97$\pm$0.18 & 3.00$\pm$0.23 & 0.48$\pm$0.20 & -- & -- & 0.71$\pm$0.32 & 1.46$\pm$0.17 \tabularnewline
CosyVoice 2 & EN & 2.46$\pm$0.26 & 3.72$\pm$0.25 & 1.41$\pm$0.41 & -- & -- & 1.80$\pm$0.50 & 2.22$\pm$0.31 \tabularnewline
ChatTTS     & EN & 2.88$\pm$0.55 & 3.31$\pm$0.38 & 3.00$\pm$0.80 & -- & -- & 3.88$\pm$0.73 & 2.94$\pm$0.74 \tabularnewline
Higgs-Audio & EN & 2.77$\pm$0.28 & 3.50$\pm$0.19 & 1.90$\pm$0.56 & -- & -- & 2.15$\pm$0.57 & 2.60$\pm$0.34 \tabularnewline
Orpheus TTS & EN & \underline{3.68$\pm$0.15} & \underline{3.76$\pm$0.12} & \underline{4.03$\pm$0.24} & -- & -- & \underline{3.97$\pm$0.26} & \underline{3.69$\pm$0.20} \tabularnewline
ElevenLabs  & EN & \textbf{3.92$\pm$0.16} & \textbf{4.21$\pm$0.13} & \textbf{4.33$\pm$0.35} & -- & -- & \textbf{4.58$\pm$0.33} & \textbf{4.12$\pm$0.23} \tabularnewline

\cmidrule(lr){1-9}

Bark        & ZH & 1.44$\pm$0.18 & 1.84$\pm$0.22 & 1.28$\pm$0.29 & -- & -- & 2.25$\pm$0.44 & 1.54$\pm$0.23 \tabularnewline
Fish-Speech & ZH & 2.65$\pm$0.26 & 3.83$\pm$0.20 & 1.29$\pm$0.37 & -- & -- & 1.58$\pm$0.42 & 2.30$\pm$0.25 \tabularnewline
Orpheus TTS & ZH & 2.81$\pm$0.22 & 3.48$\pm$0.18 & 2.74$\pm$0.33 & -- & -- & 3.09$\pm$0.34 & 2.76$\pm$0.25 \tabularnewline
CosyVoice 2 & ZH & \underline{3.37$\pm$0.16} & \underline{3.84$\pm$0.17} & 1.98$\pm$0.45 & -- & -- & 1.75$\pm$0.41 & 2.95$\pm$0.26 \tabularnewline
Higgs-Audio & ZH & 3.10$\pm$0.34 & 3.37$\pm$0.31 & \underline{2.79$\pm$0.55} & -- & -- & \underline{3.24$\pm$0.56} & \underline{3.03$\pm$0.38} \tabularnewline
ChatTTS     & ZH & 2.69$\pm$0.72 & 2.75$\pm$0.57 & 1.94$\pm$0.92 & -- & -- & 2.88$\pm$1.05 & 2.50$\pm$0.75 \tabularnewline
ElevenLabs  & ZH & \textbf{4.09$\pm$0.19} & \textbf{4.09$\pm$0.21} & \textbf{3.74$\pm$0.64} & -- & -- & \textbf{3.77$\pm$0.68} & \textbf{3.94$\pm$0.26} \tabularnewline

\bottomrule
\end{tabular}%
}

\vspace{-5pt}
\end{table*}

\vspace{-3pt}

\vspace{-4pt}
\section{Results and Analysis}
\label{sec:results}
\vspace{-4pt}

\subsection{Objective results: robust trends and NVV-specific confounders}
\label{subsec:results_objective}
\vspace{-3pt}

The objective results for both prompt-based and tag-based systems are summarized in Table~\ref{tab:prompt_tag_obj_res}. Across three independent synthesis runs, most measures exhibit small run-to-run variance, indicating that the systems are \emph{stable}.

\textbf{Prompt-based objective results.}
Results reveal a clear \emph{intelligibility--quality} split. 
In EN, \textit{Qwen3-TTS} delivers the best intelligibility and semantic alignment, achieving the lowest WER/CER and the highest CLAP score, with \textit{GPT-4o mini TTS} as a close second on CLAP. 
In contrast, \textit{GPT-4o mini TTS} provides the strongest perceived audio quality, leading the prompt-based block on DNSMOS SIG/BAK/OVRL. 
In ZH, the same pattern persists: \textit{Qwen3-TTS} attains the lowest WER/CER, while \textit{GPT-4o mini TTS} leads both DNSMOS and CLAP, indicating strong perceptual quality and caption--speech alignment.

A notable outlier is the \textit{Gemini} family, which attains competitive DNSMOS (and is also rated as natural in subjective results; Table~\ref{tab:merged_sub_res}) but shows substantially worse WER/CER, especially \textit{Gemini 2.5 Flash}. 
Qualitative inspection suggests that this inflation is mainly driven by content-boundary violations: Gemini may exhibit prompt leakage, generate hallucinated speech beyond the target text, repeat portions of the input text, or produce overly long NVV segments that ASR transcribes as repeated non-lexical tokens such as ``ha ha ha''.
These behaviors sharply increase WER/CER, even when the intended content remains understandable and the speech sounds natural. 

\textbf{Tag-based objective results.}
For tag-based systems, the results reveal a clear \textbf{breadth--correctness} trade-off. 
\textit{ChatTTS} provides the most representative case of \emph{selective compliance}: despite exhibiting the lowest coverage in both languages, it still attains highly competitive NVV-matching performance, including the best precision, recall, and F1 in ZH. This pattern suggests that ChatTTS follows tags accurately only for a very limited subset of supported cases, largely dominated by a single NVV type (e.g., \texttt{laugh}), which reduces the likelihood of type mismatch and can therefore inflate precision-, recall-, and F1-based correctness measures.

In EN, strong correctness is not concentrated in a single system. \textit{Orpheus TTS} delivers the strongest correctness profile, achieving the best precision and F1 together with competitive recall and low NTD. By contrast, \textit{ElevenLabs} offers one of the most balanced controllability profiles, combining relatively high coverage with strong correctness, while also achieving the best intelligibility among tag-based systems in EN. Taken together, these results suggest that tag-based controllability should be assessed from a coverage-aware perspective rather than inferred from any single correctness score alone.
At the other end, several open systems exhibit limited NVV-matching capability, as illustrated by \textit{Higgs-Audio}, or show relatively large NTD, as in \textit{CosyVoice 2}.
These findings indicate that favorable intelligibility or quality-related metrics do not necessarily imply reliable realization of the requested NVV type.



\textbf{Summary.}
The objective results suggest that speech generation systems with NVV generation should prioritize two concrete directions:
(i) Enforcing \textbf{content boundaries} under prompt conditioning to prevent prompt leakage, repetition, and uncontrolled extra speech.
(ii) Optimizing \textbf{coverage-aware NVV correctness} under tag conditioning to discourage selective compliance and encourage broad, reliable control across NVV types.


\vspace{-6pt}
\subsection{Subjective listening: preferences and trade-offs in NVV synthesis}
\label{subsec:results_subjective}
\vspace{-3pt}
The subjective results for prompt-based and tag-based systems are reported in Table~\ref{tab:merged_sub_res}. A total of 97 raters participated in the subjective test.
Overall, human listening reveals a consistent trade-off in NVV synthesis: 
the most preferred systems balance a natural, high-quality delivery of lexical speech with NVVs that are reliably present, correctly realized, and perceptually salient. 



\textbf{Prompt-based subjective results.}
In EN, \textit{Gemini 2.5 Pro} provides the best listening experience, achieving the top overall naturalness and quality, while also leading caption--speech match and NVV instruction following and salience. 
\textit{Gemini 2.5 Flash} remains highly competitive but is slightly less preferred overall. 
Although Gemini exhibits substantially worse WER in objective evaluation due to content-boundary violations, this does not necessarily reduce perceived naturalness, as the inserted or prolonged non-lexical segments can still sound natural and human-like.

In ZH, \textit{Qwen3-TTS} achieves the best overall naturalness and quality and also leads caption--speech match, consistent with its strong objective intelligibility and caption alignment. 
In contrast, \textit{Gemini} and \textit{GPT} use pre-defined voices that do not always reflect the voice characteristics described in the caption, which likely contributes to their lower perceptual caption--speech match scores. 
Despite this limitation, \textit{Gemini 2.5 Flash} attains the strongest NVV instruction following and NVV perceptual effect, followed by \textit{Gemini 2.5 Pro}, indicating that Gemini is particularly strong at realizing NVVs even when overall caption-level matching is imperfect. 
At the same time, the results of \textit{Gemini 2.5 Flash} suggest that more aggressive NVV realization does not always translate into higher overall naturalness. 
Our qualitative listening further shows that Flash often amplifies emphasis and prosodic variation, which can occasionally yield an unnatural delivery.

\textbf{Tag-based subjective results.}
\textit{Orpheus TTS} exhibits a particularly strong EN correctness profile in objective evaluation and remains highly competitive in EN subjective listening. By contrast, \textit{ElevenLabs} ranks best overall across languages in subjective listening tests, achieving top or near-top performance in speech naturalness and quality, together with the strongest NVV perceptual effect, correctness, and overall expressiveness. 
This suggests that objective correctness alone is insufficient to secure the best overall listening preference. The strongest systems must balance reliable NVV realization with broad coverage, naturalness, and overall audio quality.

In contrast, \textit{CosyVoice 2} highlights a characteristic failure mode: it can deliver high-fidelity speech, especially in ZH, where it attains the best perceived quality, while still lagging behind ElevenLabs on NVV correctness and salience. 
Taken together, the objective and subjective results indicate that speech fidelity and NVV controllability are partially separable dimensions that must be optimized jointly.

\textbf{Summary.}
Subjective listening suggests two practical insights for speech generation with NVV synthesis: (i)  NVVs should be inserted with well-regulated boundaries and appropriate strength, since overly aggressive realization can reduce overall naturalness. (ii)  High-fidelity speech alone is insufficient: effective NVV synthesis requires reliable tag following with broad coverage, together with perceptually correct and salient vocalization realization.

\vspace{-7pt}
\subsection{LLM-as-a-judge: scalable signals with clear limits}
\label{subsec:results_llm}
\vspace{-4pt}
LLM-based judging offers a scalable complement to human listening by enabling fast, repeatable comparisons across many systems and samples. In this work, we use Gemini 2.5 Pro as the LLM judge. As shown in Table~\ref{tab:merged_llm_res}, its ratings generally track human judgments.

\textbf{Prompt-based LLM results.}
Across both EN and ZH, \textit{Gemini 2.5 Flash} receives the strongest scores on overall instruction following and NVV-related criteria, together with the highest overall naturalness, suggesting robust instruction following from caption to speech and reliable NVV realization under natural-language descriptions. Meanwhile, \textit{Qwen3-TTS} attains the highest overall quality score in EN, echoing a recurring observation from human listening: strong perceived audio quality does not necessarily imply strong NVV controllability or semantic alignment with the caption.

\textbf{Tag-based LLM results.}
For tag-based systems, the LLM consistently favors \textit{ElevenLabs} in both EN and ZH, aligning with human ratings that it delivers strong overall speech quality with highly salient and correct NVV realization. 
Among open-source systems, relative rankings vary substantially across axes such as accuracy, expressiveness, and quality, reinforcing that NVV controllability is a partially separable dimension rather than a simple byproduct of speech fidelity.

\vspace{2pt}
\begin{figure*}[th]
  \centering
\begin{subfigure}[t]{0.42\textwidth}
    \centering
    \includegraphics[width=\linewidth,trim=8 6 8 6,clip]{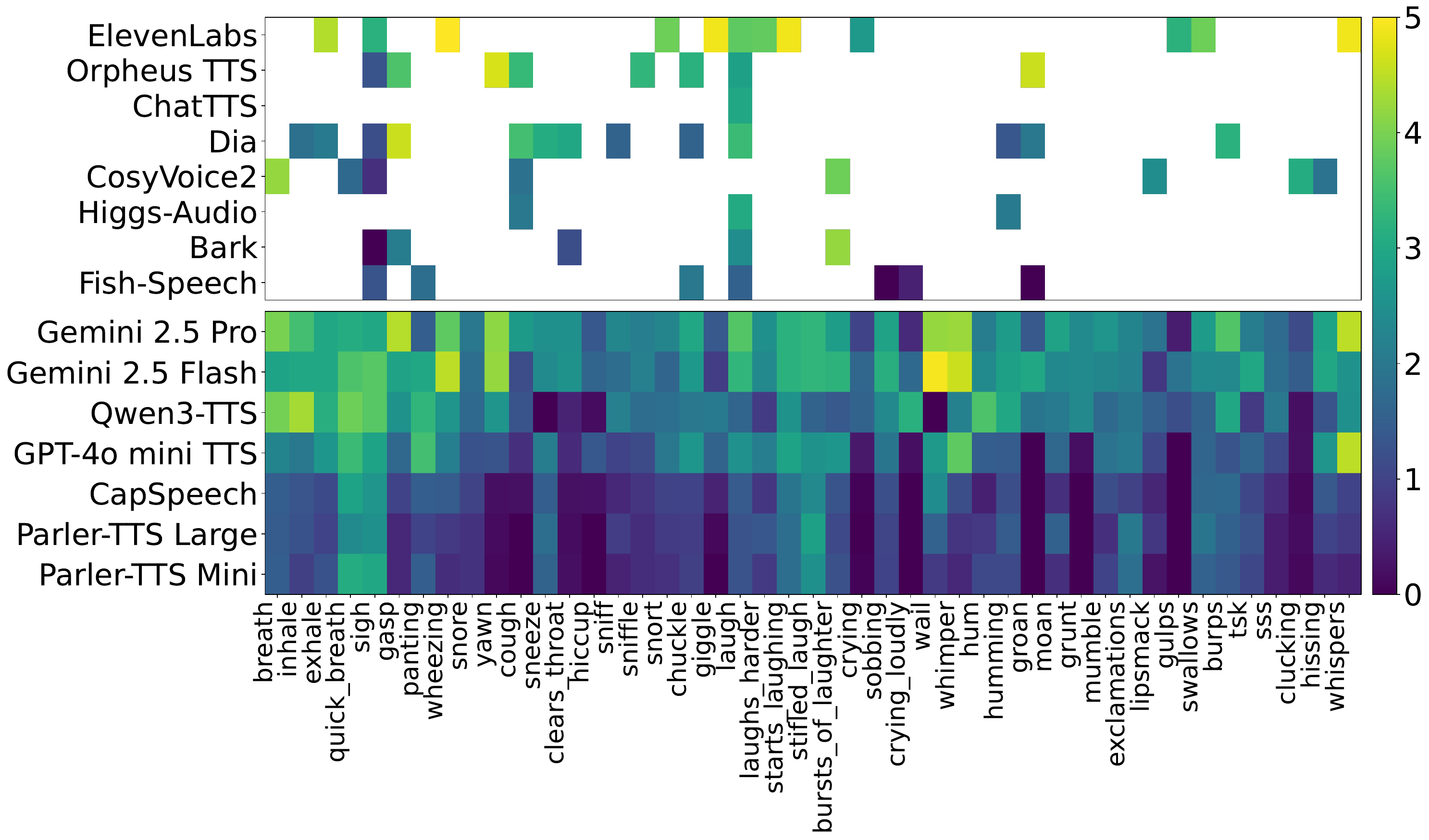}
    \vspace{-0.6em}
    \label{fig:heatmap_en_pe}
  \end{subfigure} \hfill
  \begin{subfigure}[t]{0.42\textwidth}
    \centering
    \includegraphics[width=\linewidth,trim=8 6 8 6,clip]{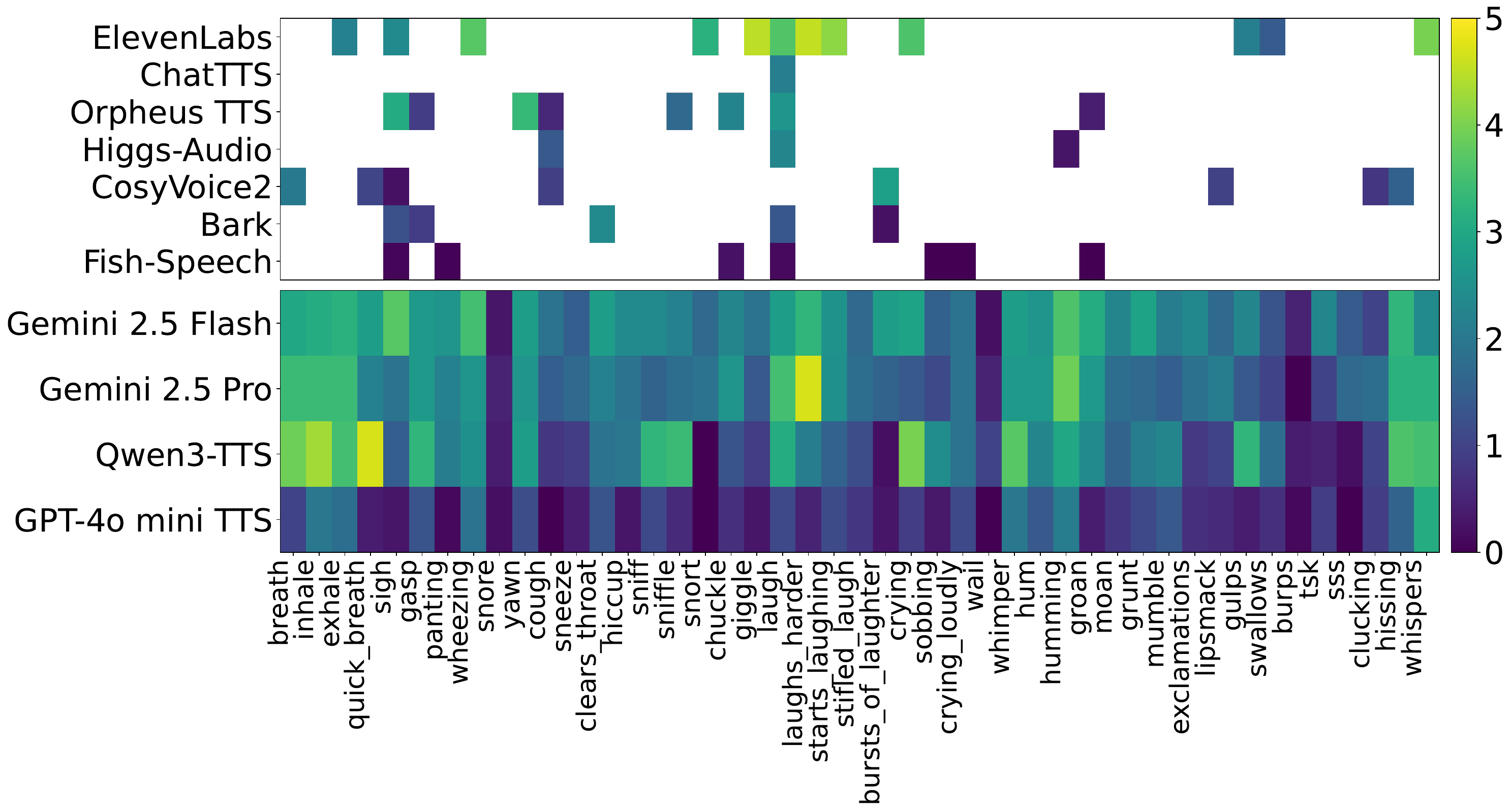}
    \vspace{-0.6em}
    \label{fig:heatmap_zh_pe}
  \end{subfigure}
  \vspace{-0.8em}
  \caption{NVV perceptual effect heatmaps for EN (left) and ZH (right) under tag-based (upper) and prompt-based (bottom) systems.}
  \label{fig:heatmap_zh_en_pe}
\vspace{-6pt}
\end{figure*}

\begin{table}[th]
  \centering
  \scriptsize
  \setlength{\tabcolsep}{3pt}
  \renewcommand{\arraystretch}{1.1}
  \vspace{-5pt}
  \caption{CMOS results for ablation study.}
  \label{tab:cmos_zh_en_two_systems}
  \begin{tabular}{llSSS}
    \toprule
    \textbf{Lang} & \textbf{System} &
    {\textbf{Naturalness}} & {\textbf{Quality}} & {\textbf{Expressiveness}} \\
    \midrule
    \multirow{2}{*}{EN} & ElevenLabs & 0.65 & 0.59 & 0.93 \\
                        & Gemini 2.5 Pro & -0.24 & -0.18 & 0.05 \\
    \midrule
    \multirow{2}{*}{ZH} & ElevenLabs & 0.33 & 0.25 & 0.52 \\
                        & Gemini 2.5 Pro & -0.14 & -0.33 & 0.05 \\
    \bottomrule
  \end{tabular}
  \vspace{-9pt}
\end{table}

\vspace{-5pt}
\subsection{Ablation: with vs.\ without explicit NVV control}
\label{sec:ablation_nvc_tag_caption}
\vspace{-3pt}
We ablate NVV synthesis by comparing paired outputs generated with versus without explicit NVV control under otherwise matched settings. We select the best-performing prompt-based system (\textbf{Gemini~2.5 Pro}) and tag-based system (\textbf{ElevenLabs}). For prompt-based speech generation, we synthesize paired samples from NVV-aware captions and their neutral counterparts, where NVV cues are removed while preserving the same propositional content. For tag-based speech generation, we synthesize paired samples from plain text and the same text with inserted NVV tags (e.g., \texttt{[sigh]}, \texttt{[laugh]}, \texttt{[sob]}). We report objective metrics for the \texttt{(w/o NVV)} condition in the gray rows of Table~\ref{tab:prompt_tag_obj_res}, and present comparative MOS (CMOS) on \textit{Naturalness}, \textit{Quality}, and \textit{Expressiveness} in Table~\ref{tab:cmos_zh_en_two_systems}. Positive CMOS indicates preference for the NVV-conditioned sample, whereas negative CMOS favors the \texttt{w/o NVV} counterpart.

\textbf{Analysis.}
Across both languages, enabling NVVs yields a consistent channel-dependent contrast between tag-based systems and prompt-based systems. On the perceptual CMOS results, \textbf{ElevenLabs} improves \textit{expressiveness}, while also increasing \textit{naturalness} and \textit{quality}. In contrast, \textbf{Gemini~2.5 Pro} gains little in \textit{expressiveness} from NVV-aware captions and tends to reduce \textit{naturalness} and \textit{quality}. The objective comparisons show higher WER/CER when NVVs are enabled, suggesting that non-lexical segments can be penalized by NVV-unaware ASR and generic quality predictors. For Gemini, this objective degradation is consistent with the CMOS drops, indicating that caption-only NVV prompting can increase generation burden without reliable perceptual benefits. For ElevenLabs, the objective drops despite positive CMOS, suggesting that standard metrics are not fully aligned with NVV-conditioned speech, motivating NVV-aware evaluation tools for non-lexical segments.

\vspace{-7pt}
\subsection{Per-type NVV perceptual effect analysis}
\label{subsec:per_type_pe}
\vspace{-4pt}
We further explore system behavior at the NVV-type level by examining \emph{per-type} perceptual salience using the subjective scores of NVV PE.  Specifically, we compute the mean NVV PE score (0--5) for each system and each NVV type, and visualize the resulting system$\times$type matrix as heatmaps in Figure~\ref{fig:heatmap_zh_en_pe}. 
In each heatmap, the tag-based systems are shown on top and the prompt-based systems on the bottom, with systems sorted within each panel by their row-mean PE. White cells denote missing entries, which primarily result from NVV types that are \emph{not supported} by tag-based inventories.

\textbf{Coverage gap.}
The tag-based panels are sparse with large white regions, consistent with the \textit{coverage} in Table~\ref{tab:prompt_tag_obj_res}: tag-based systems typically implement only $\sim$1--13 of the 45 target types (Coverage $\approx$0.02--0.29). In contrast, the prompt-based panels are dense because caption prompts are assumed to target all types.

\textbf{Type difficulty.}
Across both paradigms, high-salience events are generally easier: laughter-related cues (e.g., \texttt{laugh/laughter}), respiratory cues (e.g., \texttt{breath/inhale/exhale}), and bursty events such as \texttt{cough}, \texttt{sneeze}, \texttt{sigh}, and \texttt{gasp} tend to obtain higher PE when present. The hardest types are \textbf{low-SNR oral cues} (e.g., \texttt{tsk}, \texttt{sss}, \texttt{lipsmack}, \texttt{gulps/swallows}, \texttt{mumble}) and \textbf{long-horizon affect} (e.g., \texttt{crying/sobbing/wail/whimper}), which remain weak under prompts and are often absent from tag inventories. For tag-based systems, the heatmaps show a partial ``frequency advantage'': types that appear in many inventories (e.g., \texttt{laugh}, \texttt{cough}, \texttt{sigh}, \texttt{gasp}, \texttt{breath}) are more likely to achieve strong PE, whereas rare types (e.g., \texttt{tsk/sss}, many oral clicks/frication cues) are both less frequently supported and less reliably realized. However, frequency is not sufficient: sustained affective NVVs can be supported yet still yield only moderate PE due to coherence demands.

\textbf{System differences.}
For tag-based systems, ElevenLabs is the strongest overall system, combining relatively high coverage with consistently strong PE on common respiratory and laughter cues and support for some rarer oral types. Dia provides the broadest open-source inventory (EN-only) but shows larger type-dependent variance. Orpheus and CosyVoice 2 sit in the mid-coverage tier with uneven per-type distinctiveness. CosyVoice 2 covers several rare oral cues, but they remain difficult. ChatTTS is fundamentally limited by its minimal inventory. For prompt-based systems, Gemini 2.5 Pro and Gemini 2.5 Flash form the top tier with broadly higher PE. Qwen3-TTS is competitive but drops more on subtle cues, and GPT-4o mini TTS tends to underperform on low-SNR and long-horizon affect; open-source prompt-based systems are generally lower.

\textbf{Insights.}
NVV synthesis should be characterized by two orthogonal axes: \textbf{inventory coverage} (what can be controlled) and \textbf{per-type realization} (how salient it is once requested). For tag-based systems, expanding inventories toward underrepresented oral cues is necessary to move beyond ``easy'' NVVs. For modeling, the persistent failures point to (i) \textbf{masking-robust high-frequency detail} for low-SNR oral events and (ii) \textbf{duration and intensity trajectory control} for sustained affective NVVs. Finally, the partial link between support frequency and PE suggests that targeted data curation for rare, subtle types is likely a key driver of progress.

\vspace{-9pt}
\section{Conclusion}
\label{sec:conclusion}
\vspace{-5pt}
In this work, we introduce NVV-SuperBench, a bilingual benchmark for evaluating NVV-capable speech generation. NVV-SuperBench covers a unified 45-type NVV taxonomy and a multi-axis evaluation protocol, which separates general speech naturalness and quality from NVV controllability, placement, and perceptual salience.
We benchmark 15 speech generation systems covering both tag-based and prompt-based method via objective metrics, subjective listening tests, and LLM-based multi-rater evaluation.
The results show that NVV controllability often decouples from overall speech quality, and that subtle low-SNR oral cues and long-duration affective NVVs remain particularly difficult to synthesize. By providing a unified benchmark dataset and standardized evaluation across diverse systems and control interfaces, NVV-SuperBench lays the groundwork for improving NVV synthesis and advancing human-like speech generation.

\section{Generative AI tools}
We used large language models (LLMs) to assist three components of this work. First, LLMs were used in benchmark dataset generation to draft candidate texts and speech captions, which were then reviewed, filtered, and finalized by the authors. Second, LLMs were used as judges to support the evaluation of speech generation systems. Third, LLMs were used for limited grammar checking and polishing.  The tools used in this study include OpenAI ChatGPT and Google Gemini.

\bibliographystyle{IEEEtran}
\bibliography{main}

\end{document}